# Optical properties of gyrotropic polygonal chiral thin films

F. Babaei

Department of Physics, University of Qom, Qom, Iran
fbabaei@qom.ac.ir

Abstract

The optical response of gyrotropic and non-gyrotropic dielectric polygonal chiral thin films were investigated using transfer matrix method. The influence of gyrotropy factor, porosity and incident angle of light on Bragg regimes, selective reflection and transmission were reported. The results revealed that the Fabry-Perot oscillations were affected by gyrotropy factor and porosity of polygonal chiral thin films.

## 1. Introduction

Magneto-optical materials in a magnetic field exhibit a non-reciprocity property is called gyrotropic effect [1]. The optical characteristics of nonhomogenous and anisotropic thin films with gyrotropy property are same as plasmas and ferrites [2]. There are three kinds of gyrotropy in the names of free, forced and natural. Free gyrotropy is related to internal magnetic fields and forced gyrotropy is associated with external magnetic fields. In a natural gyrotropy, the rotation of plane of polarization is reversible (e.g. sugar solution) [1]. A gyrotropic medium can make a difference between left-handed circular polarization (LCP) and right-handed circular polarization(RCP) of light.

Polygonal chiral thin films are three dimensional artificial nanostructures as a stack of biaxial plates that can be produced by combination of oblique angle deposition technique and abrupt changes of substrate rotation about its substrate normal [3,4]. A polygonal chiral thin film can be formed with the discrete abrupt angular rotations of $2\pi/Q$, where the parameter Q>2 is an integer. A polygonal chiral thin film with Q= 4 is called equichiral sculptured thin film (or tetragonal) and other is named ambichiral sculptured thin film [5].

Chiral sculptured thin films are local biaxial [6,7] and there is a selection in reflection RCP and LCP from them in Bragg regime, which is knows as circular Bragg phenomenon (CBP) [8]. The effect of gyrotropy on the CBP exhibited by chiral sculptured thin films (Q > 10) [9,10]. In magneto-optical materials the off- diagonal components of the relative permittivity tensor can be activated (this property is called gyrotropy) [11]. Then, the mutuality between circular Bragg phenomenon and gyrotropy effect could be interested in optics of chiral magneto-optic thin films.

In this work, the problem of light propagation with circular polarization was investigated in gyrotropic polygonal chiral thin film. The effects of gyrotropy, porosity and incident angle of light on Bragg regimes was studied. The theory in brief for optical modeling is presented in Section 2 and followed by results and discussion in Section 3.

## 2. Theory in Brief

Consider the region $0 \leq z \leq d$ is occupied by a gyrotropic polygonal chiral thin film, while the regions $z \leq 0$ and $z \geq 0$ are vacuous (Fig.1). Let us assume that this structure can be exposed by a circular polarized plane wave from the bottom of the structure at an angle $\theta$ to the z- axis. The phasors of incident, reflected and transmitted electric fields are given:

$$\underline{E}_{inc}(\underline{r}) = [\frac{(i\underline{S}-\underline{P}_+)}{\sqrt{2}} a_L - \frac{(i\underline{S}+\underline{P}_+)}{\sqrt{2}} a_R] e^{ik(z\cos\theta + x\sin\theta)} , \qquad z \leq 0$$

$$\underline{E}_{ref}(\underline{r}) = [-\frac{(i\underline{S}-\underline{P}_-)}{\sqrt{2}} r_L + \frac{(i\underline{S}+\underline{P}_-)}{\sqrt{2}} r_R] e^{ik(-z\cos\theta + x\sin\theta)} , \qquad z \leq 0 \qquad (1)$$

$$\underline{E}_{tr}(\underline{r}) = [\frac{(i\underline{S}-\underline{P}_+)}{\sqrt{2}} t_L - \frac{(i\underline{S}+\underline{P}_+)}{\sqrt{2}} t_R] e^{ik((z-d)\cos\theta + x\sin\theta)} , \qquad z \geq d$$

The phasor of the magnetic field in any region is given as $\underline{H}(\underline{r}) = (i\omega\mu_0)^{-1} \nabla \times \underline{E}(\underline{r})$, where $(a_L, a_R), (r_L, r_R)$ and $(t_L, t_R)$ are the amplitudes of incident plane wave, and reflected and transmitted waves with LCP and RCP. The $\underline{S}$ and $\underline{P}_{\pm}$ are respectively unit vectors for linear polarization normal and parallel to the incident plane and also $\underline{u}_{x,y,z}$ are Cartesian unit vectors. The reflectance and transmittance amplitudes can be obtained, using the continuity of the tangential components of electrical and magnetic fields at two interfaces of structure and solving the algebraic matrix equation:

$$\begin{bmatrix} i(t_L - t_R) \\ -(t_L + t_R) \\ 0 \\ 0 \end{bmatrix} = [\underline{\underline{K}}(\theta_0)]^{-1} \cdot [\underline{\underline{M}}_P] \cdot [\underline{\underline{K}}(\theta_0)] \cdot \begin{bmatrix} i(a_L - a_R) \\ (a_L + a_R) \\ -i(r_L - r_R) \\ (r_L + r_R) \end{bmatrix} \quad (2),$$

where the different terms and parameters of this equation are given in detail in [6]. Finally, using the obtained amplitudes in Eq. 2, then we can calculate the reflectance and transmittance coefficients as $r_{ij} = \dfrac{r_j}{a_i}$ and $t_{ij} = \dfrac{t_j}{a_i}$; $i,j = L, R$. The reflection and transmission is calculated as $R_{ij} = |r_{ij}|^2$ and $T_{ij} = |t_{ij}|^2$; $i,j = L, R$.

The nonhomogeneous dielectric permittivity $\underline{\underline{\varepsilon}}_P$ for $l$ th arm of the polygonal chiral is defined as [12]:

$$\underline{\underline{\varepsilon}}_P = \underline{\underline{S}}_z(\zeta) \cdot \underline{\underline{S}}_y(\chi) \cdot \underline{\underline{\varepsilon}}_{ref} \cdot \underline{\underline{S}}_y^T(\chi) \cdot \underline{\underline{S}}_z^T(\zeta) \quad (3),$$

where the superscript T indicates to the transpose of a dyadic, $\zeta = h(l-1)(2\pi/Q)$, $\chi$ is tilt angle of the nanocolumns of the polygonal chiral, $h = +1 (or -1)$ is right-handedness (or left-handedness) of the polygonal chiral. In our work, Q was fixed as 3(trigonal),4(tetragonal),5(pentagonal),6(hexagonal) and 12(dodecagonal).

The local relative permittivity, rotation and tilt dyadics are respectively as [9,10]:

$$\begin{aligned} \underline{\underline{\varepsilon}}_{ref} &= \varepsilon_a \underline{u}_z \underline{u}_z + \varepsilon_b \underline{u}_x \underline{u}_x + \varepsilon_c \underline{u}_y \underline{u}_y + i\varepsilon_g (\underline{u}_z \underline{u}_y - \underline{u}_y \underline{u}_z) \\ \underline{\underline{S}}_z &= (\underline{u}_x \underline{u}_x + \underline{u}_y \underline{u}_y)\cos\zeta + (\underline{u}_y \underline{u}_x - \underline{u}_x \underline{u}_y)\sin\zeta + \underline{u}_z \underline{u}_z \\ \underline{\underline{S}}_y &= (\underline{u}_x \underline{u}_x + \underline{u}_z \underline{u}_z)\cos\chi + (\underline{u}_z \underline{u}_x - \underline{u}_x \underline{u}_z)\sin\chi + \underline{u}_y \underline{u}_y \end{aligned} \quad (4),$$

where $\varepsilon_{a,b,c}$ are the relative permittivity scalars and $\varepsilon_g$ indicates to gyrotropy surface of polygonal chiral in yz-plane.

## 3. Results and discussion

In order to optical modeling, at first we considered a structure as a right-handed (h= +1) $TiO_2$ porous polygonal chiral thin film (Fig.1). In optical modeling, the relative permittivity scalars $\varepsilon_{a,b,c}$ in Eq.4 are calculated using the Bruggeman homogenization formalism [13,14]. In this formalism, the structure is considered as a two-component composite ($TiO_2$ and void). These quantities are dependent on different parameters, namely columnar form factor, fraction of void ($f_v$), the wavelength of free space, and the refractive index. In addition, each column in the structure is considered as a string of identical long ellipsoids. The ellipsoids are considered to be electrically small (i.e., small in a sense that their electrical interaction can be ignored) [6]. In all calculations, columnar form factors were fixed as $(c/a)_{TiO_2, void}$ = 20 and $(b/a)_{TiO_2, void}$ =1.06 [15] (c, a & b are the semimajor axis and small half-axes of ellipsoids, respectively). We were used the frequency-dependent bulk experimental dielectric function of $TiO_2$ [16] for homogenization. The selective transmission and selective reflection is calculated as $T_{LL}$-$T_{RR}$ and $R_{LL}$-$R_{RR}$, respectively. The optical rotation is then defined as [13,14]:

$$\phi_{tr} = \begin{cases} \tau_{tr} - \tau_{inc} + \pi, & \text{if } -\pi \leq \tau_{tr} - \tau_{inc} \leq -\pi/2 \\ \tau_{tr} - \tau_{inc}, & \text{if } |\tau_{tr} - \tau_{inc}| \leq \pi/2 \\ \tau_{tr} - \tau_{inc} - \pi, & \text{if } \pi/2 \leq \tau_{tr} - \tau_{inc} \leq \pi \end{cases} \quad (5),$$

where [2]

$$\tau_{inc} = \tan^{-1}\left(\frac{|a_R|\text{Re}(a_L) + |a_L|\text{Re}(a_R)}{|a_R|\text{Im}(a_L) - |a_L|\text{Im}(a_R)}\right)$$

$$\tau_{tr} = \tan^{-1}\left(\frac{|t_R|\text{Re}(t_L) + |t_L|\text{Re}(t_R)}{|t_R|\text{Im}(t_L) - |t_L|\text{Im}(t_R)}\right) \quad (6),$$

so that Re( ) and Im ( ) are the real part and imaginary part of the quantity given in the parenthesis, respectively.

It must be mentioned here, we supposed a static magnetic field in x-direction $\underline{H} = H_0 \underline{u}_x$. In simplest case, to first order a gyration vector $g = (\varepsilon_g, 0, 0)$ is proportional to the applied magnetic field, where $\varepsilon_g = \chi^{(m)} H_0$ [11] and while $\chi^{(m)}$ is the magneto-optical susceptibility and it was a scalar and frequency independent in our simulation. In generally, it can be frequency dependent due to dispersion of $\chi^{(m)}$ and even can be a tensor. Therefore, the effect of gyrotropy in structure was considered as above not as a porous polygonal chiral thin film infiltrated with magnetic nanoparticles or a porous polygonal chiral thin film fabricated from magnetic materials. Also, we fixed the values of the level gyrotropy between -1 to +1, the azimuthal angle of incident light was 0º, the tilt angle of nanocolumns, the structural period (pitch) and the thickness of polygonal chiral thin film respectively were 42º, 2Ω=325 nm and d=3Ω.

Selective transmission from right- handed(h=+1) TiO$_2$ polygonal chiral thin films for types of polygon shapes at different void volume fractions with $\theta = 0°$ is shown in Fig.2. In our work, there exist two Bragg resonances (two Bragg regimes) due to fixed the thickness and pitch of polygonal chiral thin films, although by selection of suitable structural parameters for biaxial plates in chiral sculptured thin films can occur the multiple Bragg regimes [3]. The first Bragg resonance(BR) centered around at wavelength $\lambda^{BR} \approx 2n_{avg} \Omega$, where $n_{avg}$ is average refractive index of polygonal chiral thin film and the second Bragg resonance almost located at wavelength $(2/(Q-2)) \lambda^{BR}$. It is well known that when $\varepsilon_g = 0$, the average refractive index of porous polygonal chiral thin is $n_{avg} = \dfrac{\sqrt{\varepsilon_c} + \sqrt{\varepsilon_d}}{2} \cos^{1/2} \theta$, where $\varepsilon_d = \dfrac{\varepsilon_a \varepsilon_b}{\varepsilon_a \cos^2 \chi + \varepsilon_b \sin^2 \chi}$ is composite relative permittivity scalar[6].If we consider $\varepsilon_g \neq 0$, the relative permittivity scalars $\varepsilon_{a,b,c}$ are modified as $\varepsilon_b \to \varepsilon_b, \varepsilon_c \to \dfrac{(\varepsilon_c + \varepsilon_a) + \sqrt{(\varepsilon_c - \varepsilon_a)^2 + 4\varepsilon_g^2}}{2}$ and $\varepsilon_a \to \dfrac{(\varepsilon_c + \varepsilon_a) - \sqrt{(\varepsilon_c - \varepsilon_a)^2 + 4\varepsilon_g^2}}{2}$. These new relative permittivity scalars can be obtained from diagonalization the local relative permittivity in Eq.4. It is enough that the new relative permittivity scalars to be replaced in definition of the average refractive index. The wavelength of first and second circular Bragg resonances (BR) extract from selective transmission are given in Fig.3. It is found that the intensity and width of Bragg regimes increased by increasing the porosity of polygonal chiral thin films from 0.2 to 0.8. On the other hand, increase porosity of polygonal chiral thin film enhances the circular Bragg phenomenon (in intensity and width). In all types of polygon shapes, the Bragg regimes shift to shorter wavelengths with increase in porosity. It can be seen that the optical properties of polygonal chiral thin films are same as the chiral sculptured thin films at polygon shapes with more vertices. For Q = 4(tetragonal), two Bragg resonances coincide at wavelength in $\lambda^{BR}$ for different porosity of polygonal chiral thin films. Also, the calculations were performed for oblique incident angle of light and obtained same results. Here, only for Q=3(trigonal shape) density plots as functions of wavelength and incident angle[0º-45º] at different void volume fractions depicted in Fig.4 so that other parameters are same as Fig.2. It is clear that in Fig.4 the both Bragg regimes move to shorter wavelengths with increasing porosity of structure.

Reflection & transmission, selected reflection & transmission of non-gyrotropic and gyrotropic ($\varepsilon_g$ =0.05) hexagonal chiral thin film(Q=6) and their optical rotation at $\varepsilon_g$ = -0.05,0.0,0.05 with $f_v$=0.4 are given in Fig.5. The obtained results showed that the gyrotropy property had not significant effect on co- and cross-handed reflection and transmission spectra. The oscillations outside of direct Bragg regime in selective transmission can be related to the Fabry-Perot oscillations [17]. It is clear that the gyrotropy property can change the Fabry-Perot oscillations due to variations of refractive index of polygonal chiral thin film.

Optical rotation spectra revealed that the gyrotropy property is an asymmetric parameter (see the spectra for $\varepsilon_g$ = -0.05 and $\varepsilon_g$ = 0.05 in Fig.5) and also the values of optical rotation significantly shift in outside of first Bragg regime to its values in non-gyrotropic polygonal chiral thin film.

The selected reflection & transmission of gyrotropic ($\varepsilon_g$ =0.1) polygonal chiral thin films for types of polygon shapes with $f_v$=0.4 and selected reflection & transmission of gyrotropic polygonal chiral thin film(Q=3) with $\varepsilon_g$ =0.1 with different porosity are shown respectively in Fig.6 and Fig. 7. It is found that the selective transmission spectra had oscillations outside of Bragg regimes in gyrotropic polygonal chiral thin film for types of polygon shapes, while these oscillations do not exist in Bragg regimes (see Fig.6) and also these oscillations disappeared in selective reflection spectra. A comparison in selective transmission spectra between Fig.2 and Fig.6 showed these oscillations do not exist in non-gyrotropic polygonal chiral thin film. Then it can be concluded that the gyrotropy property created them in polygonal chiral thin films due to activation the off- diagonal components of the relative permittivity tensor. It can be observed in Fig.7 for trigonal chiral thin film that the oscillations outside of Bragg regimes decreased with increasing porosity of polygonal chiral thin films from 0.2 to 0.8. Hence, it can be expressed that the gyrotropy does not affect on oscillations outside of Bragg regimes in selective transmission spectra at high porosity.

In Fig.8, selected transmission & reflection of gyrotropic polygonal chiral thin film (Q=3) at different level of gyrotropy with fv=0.4 , the wavelength of two Bragg resonances and the intensity of selective reflection (in percent) at wavelength of two Bragg resonances are shown. The obtained results showed that the location of occurrence of first Bragg resonance was almost constant (630nm) by variations of level of gyrotropy and it is obvious from the selective transmission and reflection spectra. But there are slight changes around 1200nm in second Bragg resonance. It is seen that these oscillations exist at different level of gyrotropy in selective transmission spectra. The intensity of selective reflection spectra at first and second Bragg resonances showed slight changes around -80% and 80%, respectively.

Density plots of selective transmission and reflection of trigonal chiral thin film (Q=3) for normal incident of light(θ=0º) as functions of wavelength and gyrotropy factor at different porosity are depicted in Fig.9.The obtained results showed that both of Bragg regimes shift to shorter wavelengths with increasing porosity of polygonal chiral thin film. Oscillations outside of Bragg regimes in density plots of selective transmission still exist due to variations of refractive index of polygonal chiral thin film. It is found that the intensity of Fabry-Perot oscillations decreased in selective transmission with increasing of porosity of polygonal chiral thin film. Because the optical constants of structure close to vacuous medium. The calculations were repeated for oblique incident angles of light with fv=0.4 that they are given in Fig.10. The obtained results are same as Fig.9. The only difference was between selective transmission and reflection spectra in direct and inverse Bragg regimes. The chiral sculptured thin films almost completely reflect the circular polarization state of normally incident light in the Bragg regime when it matches the structural handedness of the material, while the opposite circular polarization state can freely propagate[6].A polygonal thin film can be described as a stack of biaxial plates that rotate by 2π/Q increments in the clockwise direction as seen by RCP light and by π(Q-2)/Q increments in the counterclockwise direction as seen by LCP light[3,4]; these different rotation rates are responsible for the difference Bragg regimes in transmission and reflection spectra .Obtained results in Fig.9 and Fig. 10 showed that the width of Bragg resonances in density plots are symmetric to the $\varepsilon_g$ =0.0 for polygonal chiral thin film at θ=0°, while it is antisymmetric at oblique incident of light.

## 4. Conclusions

The optical properties of gyrotropic polygonal chiral thin films were investigated using transfer matrix method. It was observed that the intensity of optical rotation spectra in a gyrotropic polygonal chiral thin film at all levels of gyrotropy was higher than a non-gyrotropic polygonal chiral thin film. The polygon shapes of gyrotropic polygonal chiral thin films do not effect from gyrotropy property and they involved in creation of circular Bragg phenomenon. We found that the intensity and width of Bragg regimes increased by increasing the porosity of polygonal chiral thin films from 0.2 to 0.8. The Fabry-Perot oscillations (out of Bragg regimes) in selective transmission spectra increased with rising the level of gyrotropy. First and second Bragg resonances shift to blue wavelengths with increasing the level of gyrotropy and incident angle of light. The width of Bragg resonances was symmetric to the axis $\varepsilon_g$ =0.0 in polygonal chiral thin film for normal incident angle of light, while there is not a symmetric at oblique incident angle of light.

**Acknowledgements**

This work was carried out with the support of the University of Qom. The author thanks the reviewers for good comments.

**References**
[1] A.D. Boardman, M. Xie: Magneto-optics: A critical review. In: W.S. Weiglhofer , A. Lakhtakia (eds.): Introduction to Complex Mediums for Optics and Electromagnetics. SPIE Press (2003) Bellingham, WA, USA.
[2] H.C. Chen: Theory of Electromagnetic Waves. McGraw-Hill (1983) New York,USA.
[3] I. J. Hodgkinson, A. Lakhtakia, Q. H. Wu, L.D.Silva, M.W. McCall, Ambichiral, equichiral and finely chiral layered structures,Opt. Commun 239(2004) 353–358.
[4] A.C.van Popta, M.J. Brett, J.C. Sit, Double-handed circular Bragg phenomena in polygonal helix thin films, J. Appl. Phys.98( 2005)083517.
[5] T.G.Mackay, J.A. Polo Jr, A. Lakhtakia ,Electromagnetic surface waves: a modern perspective. Elsevier (2013) Waltham.
[6] A. Lakhtakia , R. Messier, Sculptured thin films: Nanoengineered Morphology and Optics (SPIE, 2005).
[7] A. Lakhtakia, On percolation and circular Bragg phenomenon in metallic helicoidal periodic ,sculptured thin films,Microw. Opt. Technol. Lett. 24 (2000) 239-244.
[8] M.Faryad,A.Lakhtakia, Advances in Optics and Photonics 6 (2014)225-292.
[9] M.D. Pickett, A. Lakhtakia, J.A. Polo Jr, Spectral responses of gyrotropic chiral sculptured thin films to obliquely incident plane waves, Optik.115(2004)393-398.
[10] M.D. Pickett, A. Lakhtakia, On gyrotropic chiral sculptured thin films for magneto- optics, Optik.113(2002)367-371.

[11] A. Eroglu, Wave propagation and radiation in gyrotropic and anisotropic media, Springer(2010) New York ,USA.
[12] T. G. Mackay, J.A.Polo, and A. Lakhtakia, Electromagnetic Surface Waves: A Modern Perspective (Elsevier, 2013).
[13] J.A.Sherwin, A.Lakhtakia, Nominal model for the optical response of a chiral sculptured thin film infiltrated with an isotropic chiral fluid, Opt Commun 214(2002)231–245.
[14] J.A.Sherwin, A.Lakhtakia, Nominal model for the optical response of a chiral sculptured thin film infiltrated by an isotropic chiral fluid-oblique incidence, Opt Commun 222(2003)305–329.
[15]J.A.Sherwin, A.Lakhtakia, I.L.Hodgkinson , On calibration of a nominal structure–property relationship model for chiral sculptured thin films by axial transmittance measurements, Opt Commun 209(2002)369–375.
[16] E.D. Palik , Handbook of optical constants of solids. Academic(1985),New York.
[17] N.P.Mahalik, Micromanufacturing and Nanotechnology. Springer (2006), Berlin, Heidelberg.

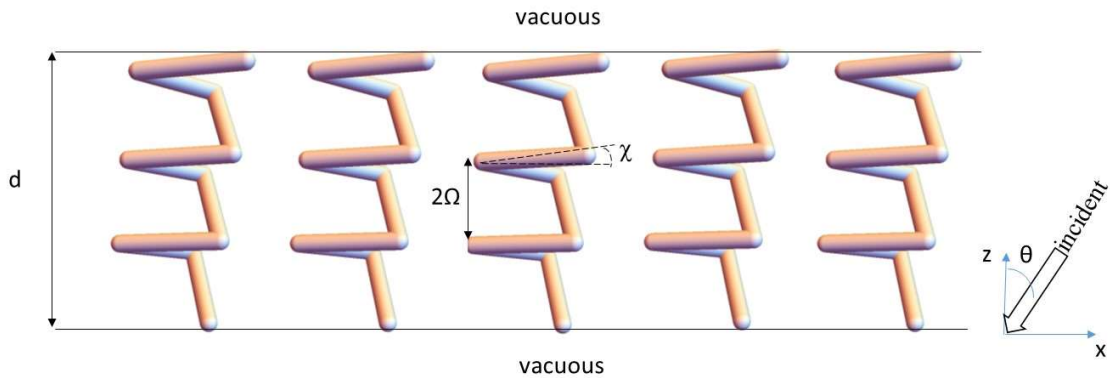

**Fig. 1.** A schematic of boundary value problem for polygonal chiral thin film (case Q=3, trigonal) in vacuous medium. The incident angle of light is $\theta$, tilt angle of nanocolumns is $\chi$ and $2\Omega$ is pitch of polygonal chiral thin film.

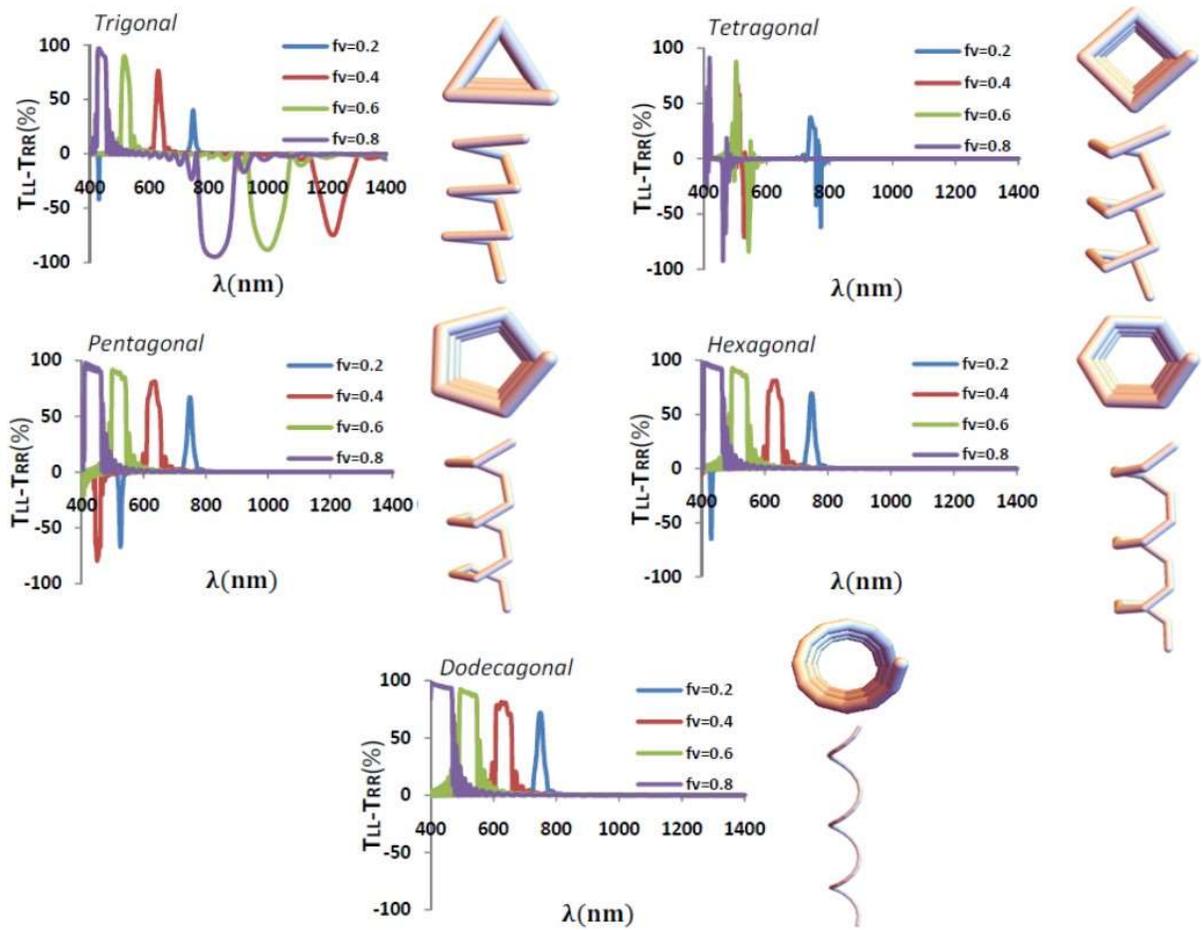

**Fig.2.** Selective transmission from right- handed(h=+1) non-gyrotropic $TiO_2$ polygonal chiral thin films at different void volume fractions with $\theta = 0°$, $2\Omega = 325\,nm$, $d_{RP} = 3\Omega$ and $\chi = 42°$. Inset plots show a single column of polygonal chiral thin films and its cross section.

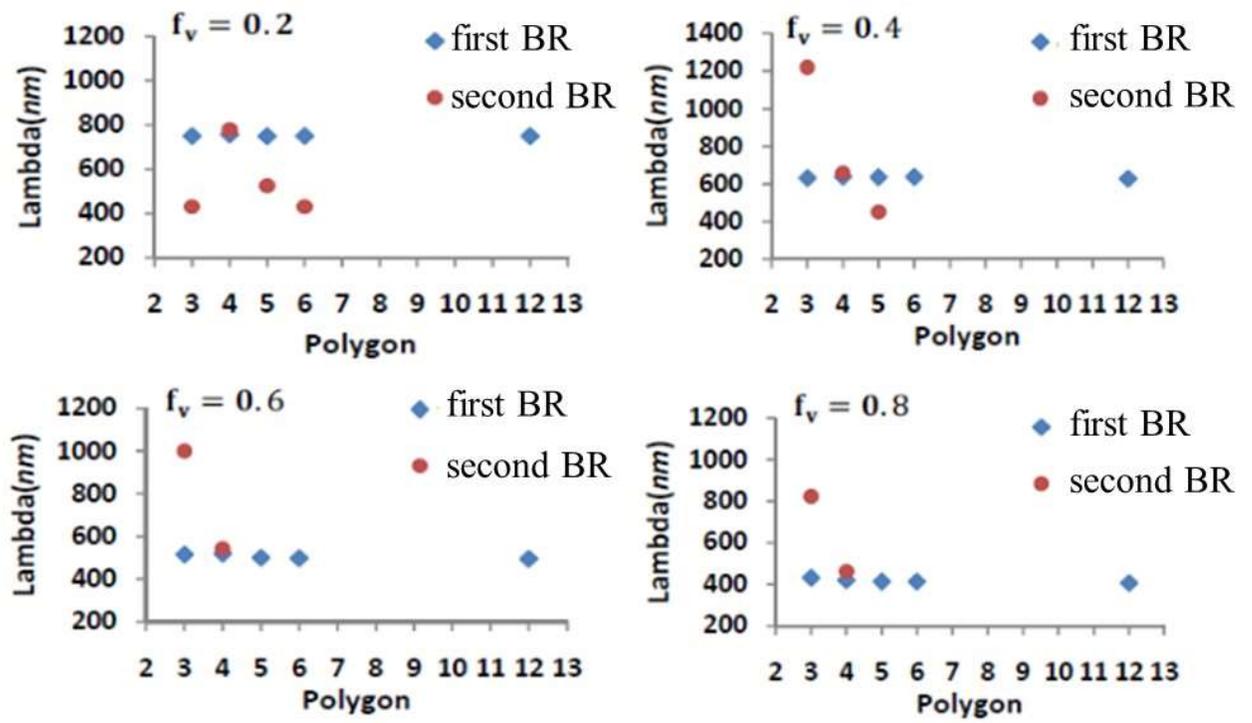

**Fig.3.** The wavelength of first Bragg resonance and second Bragg resonance for different polygonal chiral thin film.

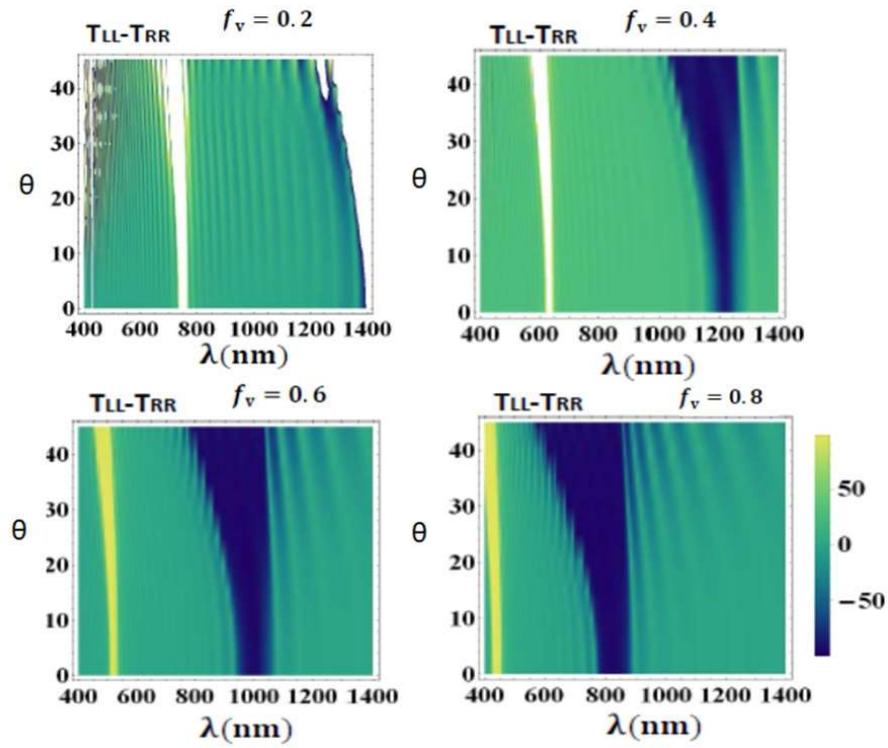

**Fig.4.** Density plots of right- handed non-gyrotropic TiO$_2$ trigonal chiral (Q=3) as functions of wavelength and incident angle at different void volume fractions.

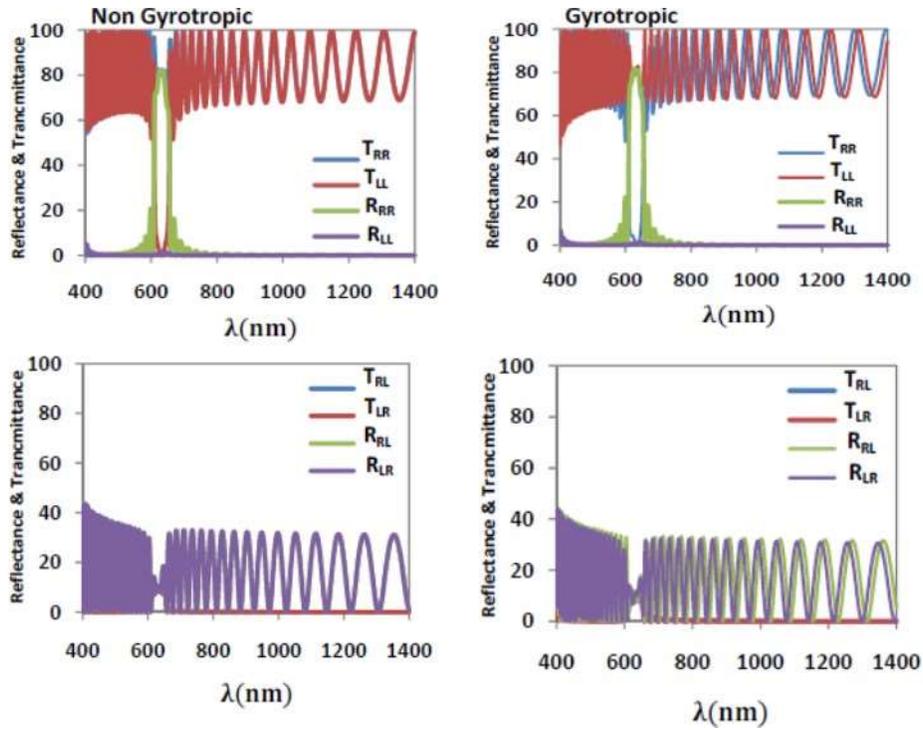

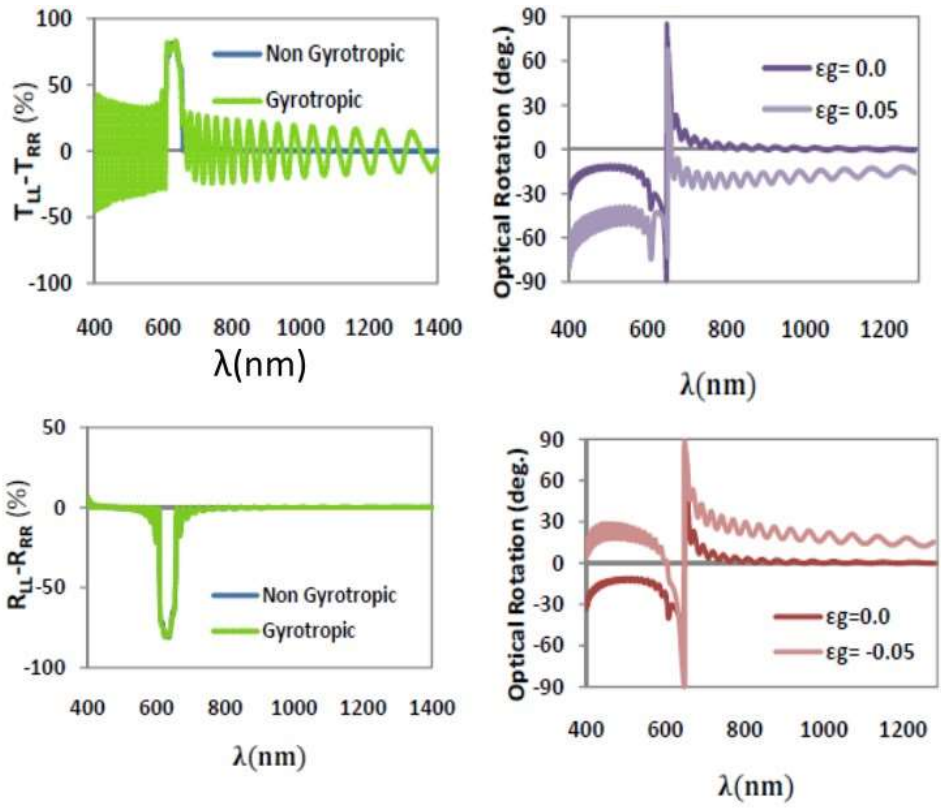

**Fig. 5.** Reflection & transmission, selected reflection& transmission of non-gyrotropic and gyrotropic ($\varepsilon_g$=0.05) hexagonal chiral thin film (Q=6) and their optical rotation at $\varepsilon_g$ = -0.05,0.0,0.05 with $f_v$=0.4.

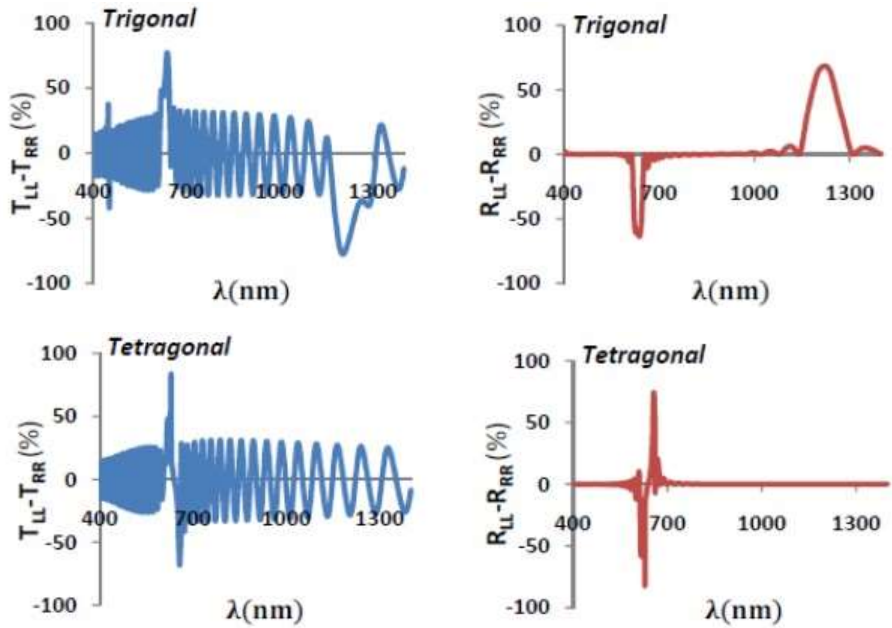

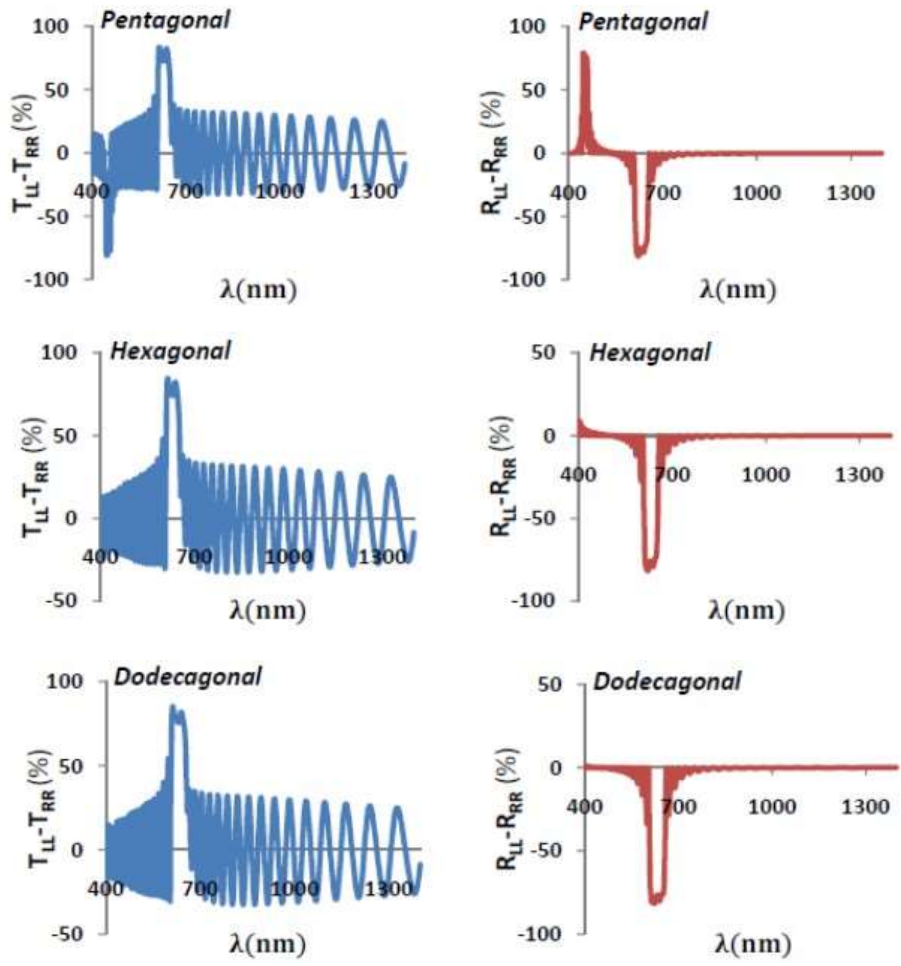

**Fig. 6.** Selected reflection & transmission of gyrotropic ($\varepsilon_g$ =0.1) polygonal chiral thin films with $f_v$=0.4.

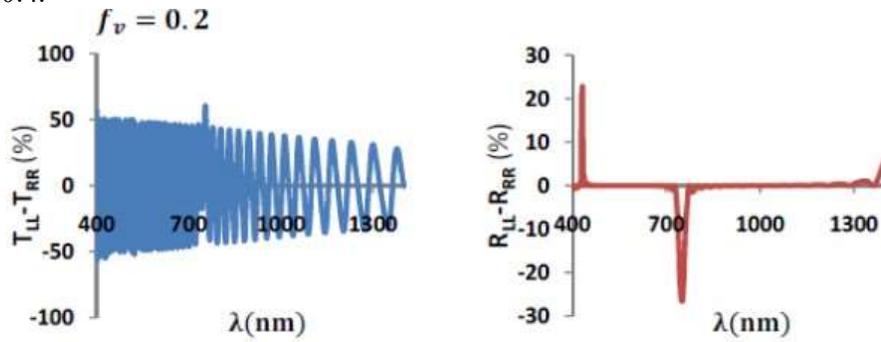

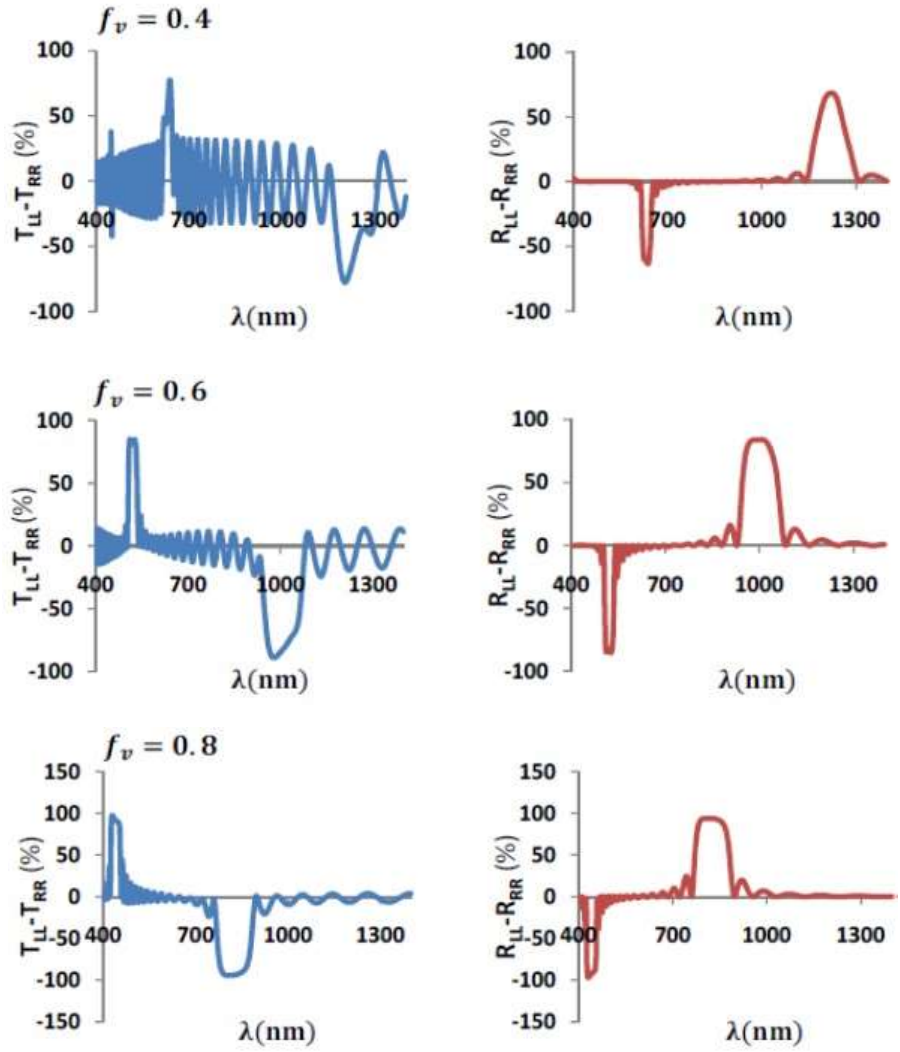

**Fig.7.** Selected reflection& transmission of gyrotropic trigonal chiral thin film (Q=3) with $\varepsilon_g$ =0.1 with different porosity.

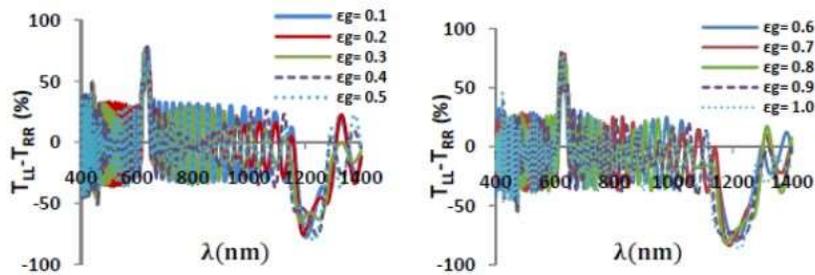

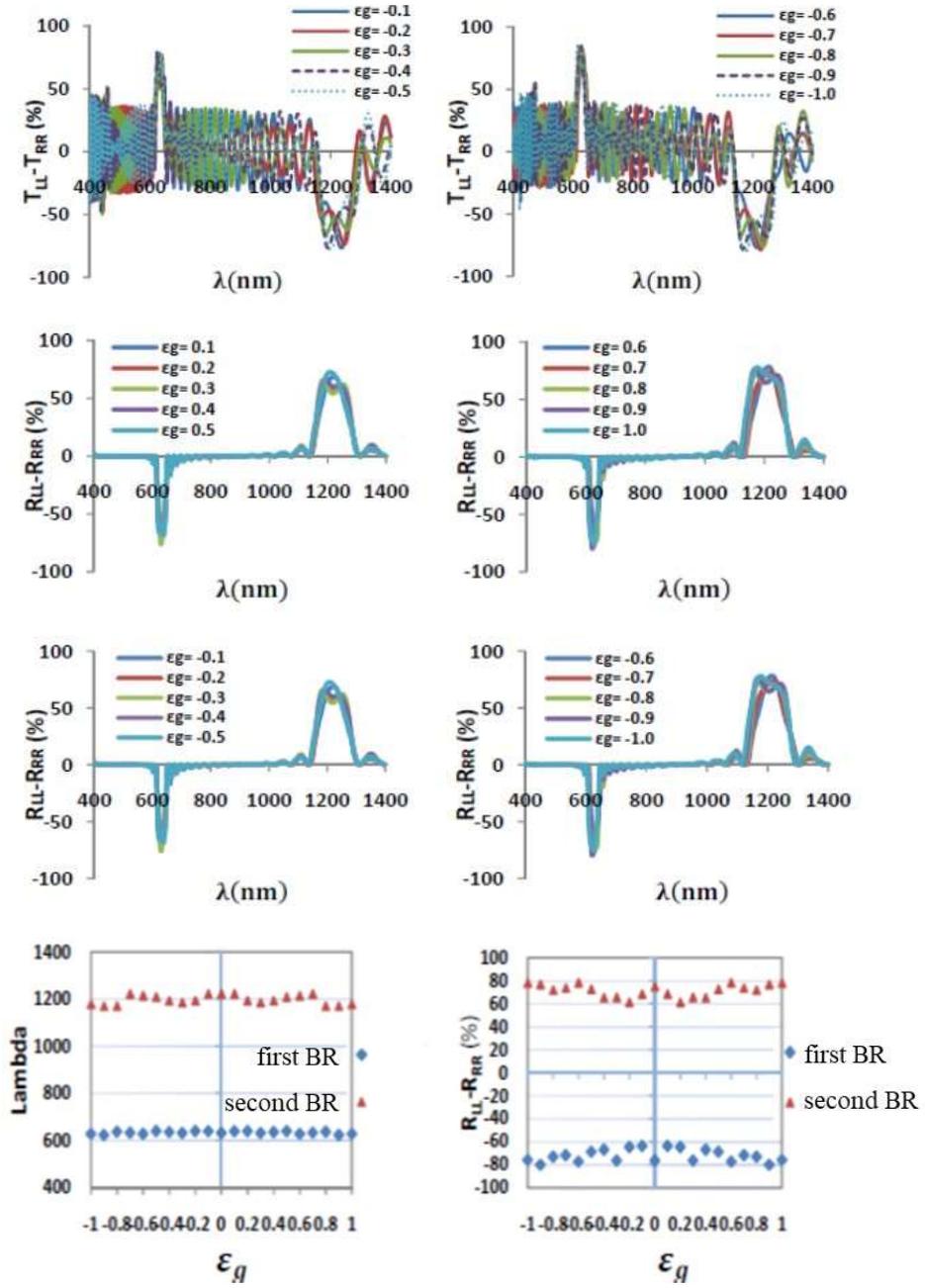

**Fig. 8.** Selected transmission and reflection of gyrotropic trigonal chiral (Q=3) at different $\varepsilon_g$ with fv=0.4. The wavelength of first and second Bragg resonances and also the intensity of selected reflection (in percent) at two Bragg resonances is given at different $\varepsilon_g$.

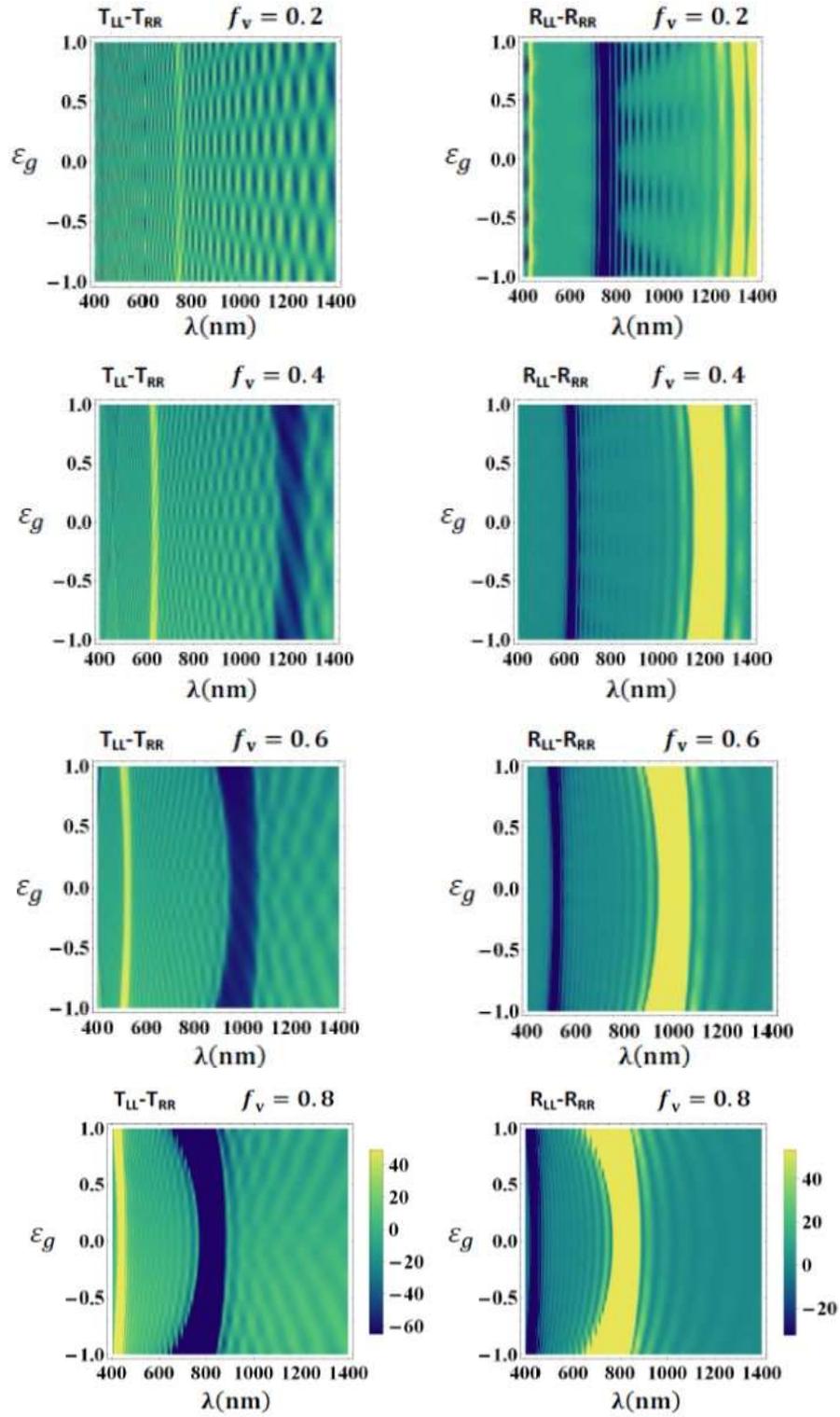

**Fig. 9.** Density plots of selected reflection and transmission of trigonal chiral thin film (Q=3) as functions of $\lambda$ and $\varepsilon_g$ at different void volume fractions for normal incident of light($\theta=0°$).

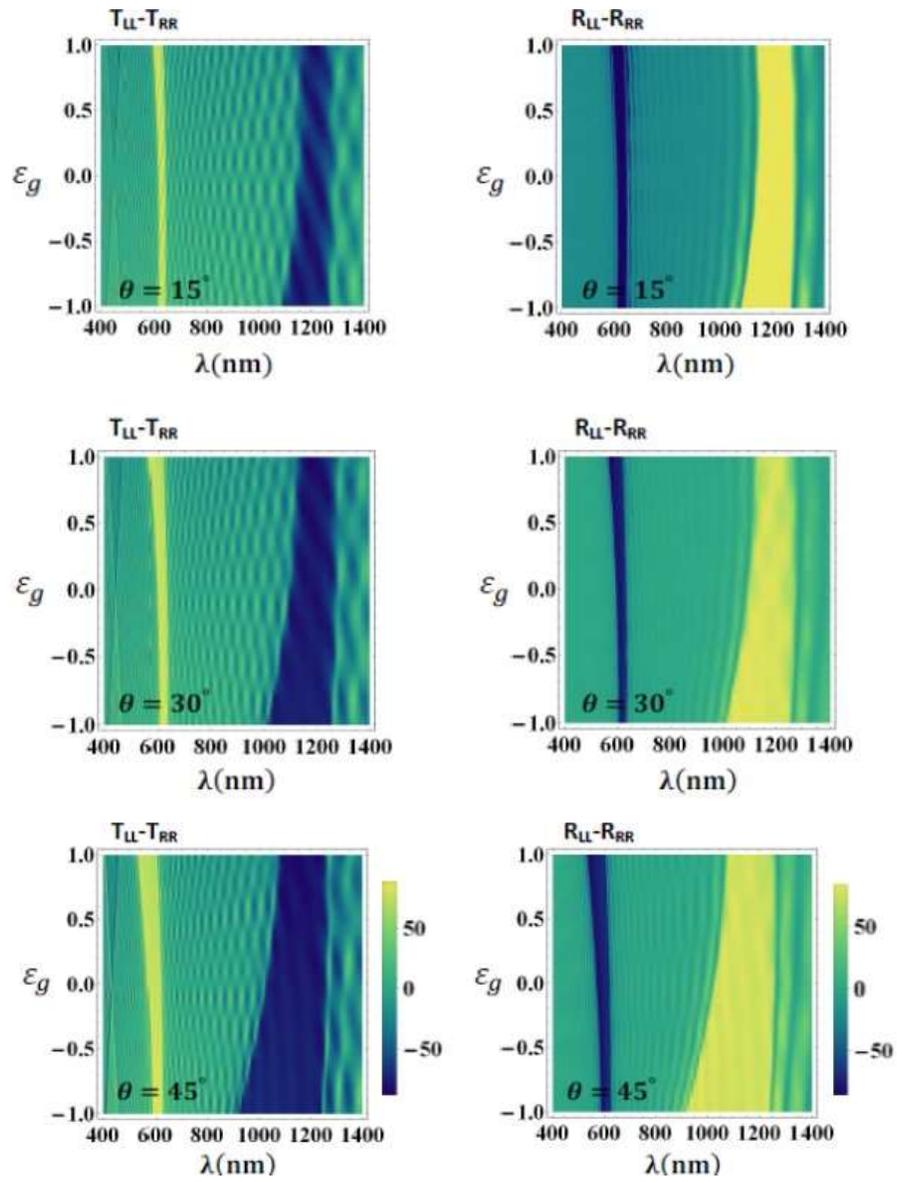

**Fig. 10.** Density plots of selected reflection and transmission of trigonal chiral thin film (Q=3) as functions of $\lambda$ and $\varepsilon_g$ at different incident angles with $f_v$=0.4.